# Knowledge Compression via Question Generation: Enhancing Multihop Document Retrieval without Fine-tuning

**Anvi Alex Eponon**[*], **Moein Shahiki-Tash**[*], **Ildar Batyrshin**,
**Christian E. Maldonado-Sifuentes**, **Grigori Sidorov**, **Alexander Gelbukh**
Instituto Politécnico Nacional (IPN), Centro de Investigación en Computación (CIC),
Mexico City, Mexico
{aeponon2023, mshahikit2022, batyr1, cmaldonado2023, sidorov, gelbukh}@cic.ipn.mx

**Code:** https://github.com/anvix9/llama2-chat/

## Abstract

This study presents a question-based knowledge encoding approach that enhances the performance of Large Language Models in Retrieval-Augmented Generation systems without requiring model fine-tuning or traditional chunking strategies. We encode textual content through generated questions that span the information space lexically and semantically, creating targeted retrieval cues paired with a custom syntactic reranking method.

Experiments on a single-hop retrieval task using 109 scientific papers show our approach achieves a Recall@3 of 0.84, outperforming traditional chunking methods by 60%. We introduce "paper-cards"—concise paper summaries under 300 characters—that significantly improve BM25 performance, increasing MRR@3 from 0.56 to 0.85 on simplified technical queries.

For multihop tasks, our syntactic reranking method reaches a 0.52 F1-score with LLaMA2-Chat-7B on the LongBench QA v1 2WikiMultihopQA dataset, surpassing chunking and fine-tuned baselines scoring 0.328 and 0.412.

This method eliminates fine-tuning requirements, reduces retrieval latency, enables intuitive question-driven knowledge access, and decreases vector storage demands by 80%, positioning it as a scalable and efficient RAG alternative.

## 1 Introduction

Retrieval-augmented generation (RAG) systems have emerged as a practical solution to one of the most pressing issues in modern Large Language Models (LLMs): hallucination. By grounding generation in external knowledge sources, RAG systems significantly reduce the likelihood of generating false or misleading information (Zhu et al., 2024). This integration enables LLMs to incorporate real-world, up-to-date, data into their outputs. However, the nature of LLMs, optimized for handling large volumes of data, introduces new challenges, especially when scaling external information. As the size of the document corpus grows, the difficulty of efficiently retrieving relevant information increases, often affecting the overall performance and response quality.

To address this, research has largely focused on architectural innovations and domain-specific fine-tuning strategies aimed at enhancing semantic understanding. While these directions are valuable, one critical but often under-emphasized component is the chunking strategy, the method by which documents are segmented for retrieval. Chunking plays a pivotal role in determining retrieval accuracy. Even the most semantically aware models can underperform if the input context is poorly segmented or misaligned with the query intent.

In this study, we propose a novel approach that leverages question and query generation as a form of **knowledge compression** to enable more effective single and multi-document retrieval. The experiments conducted aim at answering:

- How can question-based knowledge encoding improve retrieval performance in RAG systems compared to traditional chunking methods?

- What is the impact of syntactic reranking and "paper-card" summaries on the accuracy and efficiency of information retrieval in scientific texts?

- Can question generation serve as an effective form of knowledge compression for scalable, fine-tuning-free RAG architectures?

In the following sections we introduce a detailed literature review, we present the motivations for our research, outline our methodology, and discuss experimental results that benchmark our method against existing approaches.

## 2 Literature Review

Tasks such as document retrieval and information extraction in systems like RAGs (Retrieval-Augmented Generation) have been of sustained interest within the research community. This is because it continually presents challenges whose solutions deepen our understanding of intelligence (mostly about retrieval processes) and its application through machines.

One of the core challenges in improving RAG systems lies in the chunking mechanism. Chunking refers to the strategic splitting of text into segments that preserve complete and meaningful information, enabling accurate retrieval in response to a query (Xu et al., 2023). Several established chunking strategies have been adopted in RAG systems. These include fixed-size chunking, structure-based (or recursive) chunking , semantic or contextual chunking (e.g., clustering based on embedding similarity), and hybrid chunking (a combination of multiple techniques) (Jimeno-Yepes et al., 2024). Among these, semantic chunking has received particular attention, especially with the emergence of fine-tuned models specifically designed to perform chunking more precisely semantically speaking.

Fixed-size chunking was the first widely adopted technique, emerging with the rise of RAG systems around 2020. It involves dividing a document into equally sized spans, often with an overlapping window to help retain context across segments. When implemented thoughtfully, this method has proven to be both robust and reliable. For instance, (Lewis et al., 2020) introduced RAGs using fixed-length passages of 100 words. Later, (Qu et al., 2025) compared fixed-size and semantic chunking techniques and found that a fixed size of 200 words could match or even outperform semantic chunking on real-world datasets, also supported by (Merola and Singh, 2025). Their findings highlighted that fixed-size chunking remains a competitive option, especially considering its lower computational cost relative to semantic approaches.

Structure-based (or recursive) chunking leverages the inherent hierarchical organization of documents, such as headings, sections, and tables, to determine natural segmentation boundaries. This approach eliminates the need to specify arbitrary token counts for each chunk. (Jimeno-Yepes et al., 2024) evaluated structure-based chunking on the FinanceBench QA dataset and demonstrated that it consistently outperformed fixed-size chunking across various chunk lengths (128-512 words). Their optimal retrieval configuration achieved 84% accuracy with ROUGE and BLEU scores of 0.57 and 0.45, respectively. Building on this foundation, (Gong et al., 2020) developed a recurrent chunking mechanism to overcome the token limitations of models like BERT when processing extended texts. Their approach employs reinforcement learning to dynamically determine segment boundaries and facilitate information flow across segments. When evaluated on the CoQA and QuAC datasets, their method surpassed BERT-large (Devlin et al., 2019), achieving F1 scores of 81.8 and 62.0, respectively.

More recently, Zhao in (Zhao et al., 2024) introduced "Meta-Chunking", which segments text into units that are larger than sentences but smaller than paragraphs, based on logical linguistic connections. Using variants of Qwen2 and Baichuan2 models, they demonstrated superior performance compared to traditional chunking approaches on both BLEU/ROUGE metrics and F1 scores on the LongBench 2WikiMultihopQA dataset.

Despite their advantages, structure-based approaches face significant limitations in real-world applications. A major challenge is that many documents found online lack clean, consistent structural formatting. Web content often combines irregular layout elements such as embedded multimedia (images, audio, video), dynamic content, or non-standard HTML/CSS structures, which makes automated structural parsing difficult and unreliable. This variability reduces the effectiveness of structure-based chunking in practical, web-scale scenarios.

Semantic chunking represents the most actively explored direction in current research. This approach typically involves computing dense or sparse embeddings and clustering text based on semantic similarity, allowing related content to be grouped regardless of its distribution throughout the document. However, as it is observed in (Qu et al., 2025), basic semantic chunking only occasionally enhances retrieval performance while substantially increasing computational demands. In many implementations, it serves primarily as a baseline for comparison with more sophisticated techniques, as demonstrated by (Zhao et al., 2025), who showed that LLM-based or learned chunkers generally produce superior results.

To address these limitations, strategies have been developed to improve chunking in RAG systems for more effective and relevant retrieval.

LumberChunker by Duarte et al. in (Duarte et al., 2024) presents a novel approach that leverages LLMs interactively to segment long narrative texts. The system begins with an initial window of passages and prompts the language model to identify natural content transitions that indicate appropriate chunk boundaries. When integrated into an RAG pipeline, this method achieved higher QA accuracy than both fixed-size and simpler chunking approaches, even matching the performance of Gemini 1.5 Pro in some contexts.

Another significant contribution is Chunking-Free Retrieval by (Qian et al., 2024), which eliminates the need for predefined chunks altogether. Their model, CFIC (Chunk-Free In-Context), encodes the entire document with a transformer and uses in-context generation to extract evidence directly. It generates token prefixes aligned with sentence boundaries via constrained decoding, thereby producing coherent supporting spans. On the LongBench QA benchmark, CFIC achieved superior F1 scores using LLama2-Chat-7B and Vicuna1.5-7B-16k compared to traditional chunking-based approaches.

Among the more sophisticated methods, Mixture of Chunkers (MoC) by (Zhao et al., 2025) introduces a routing mechanism that dynamically selects from a set of "meta-chunkers", including rule-based, regex-guided, and LLM-based chunkers. MoC's router learns which chunking expert to apply based on the context. Additionally, (Zhao et al., 2025) propose intrinsic metrics such as Boundary Clarity and Chunk Stickiness to evaluate chunk quality beyond retrieval accuracy. Their method consistently outperformed other models and techniques, such as LlamaIndex and standard semantic chunking, in chunking-related tasks.

An interesting approach that supports our current study is the experiment conducted by Zackary R. on Rag-Fusion: A New Take on Retrieval Augmented Generation (Rackauckas, 2024), where the author generated multiple queries and reranking them with reciprocal scores and fusing the documents and scores. The approach presents a good advantage in the sense that it provides more context to the model during the request (online) while displaying more disadvantages such as long processing time. Since query generation occurs during the request phase, users experience longer waiting periods, and the absence of targeted query generation can result in off-topic queries that diminish retrieval effectiveness.

While these techniques mark significant progress in intelligent and adaptive chunking, they also bring a set of new challenges:

- **High computational costs:** Many advanced chunking strategies rely on large language models or transformer encoders, which demand considerable computational resources, especially during fine-tuning and inference.

- **Need for Domain-specific fine-tuning:** Several methods, such as MoC and CFIC, require fine-tuning of models or training of auxiliary components like routers, increasing implementation complexity and limiting accessibility for low-resource settings.

- **Latency and scalability:** Interactive or dynamic chunking methods, while more precise, may introduce latency that hampers real-time retrieval applications or large-scale deployment.

Based on the identified limitations, our research aims to develop an approach that enhances chunking and retrieval processes without requiring model fine-tuning.

- The proposed approach combines explicit syntactic reranking with the semantic understanding capabilities of existing models for retrieval tasks, involving the generation of both queries and predefined questions that function as semantic boundaries within the text.

- Then, from generated questions, queries and keywords, a compact paper-card is created which represents the main information of the paper as an alternative and more rich version of a traditional abstract.

- We evaluate the approach against traditional chunking methods (fixed-size, structure-based, BM25) and assess its performance on the LongBench QA v1 (Bai et al., 2024) dataset, specifically using the 2WikiMQA and the 2WikiMQA_e datasets. The evaluation compares results across multiple models, including LLaMA3, LLaMA2-chat-7B (Touvron et al., 2023), and Vicuna-7B (Zheng et al., 2023), GPT-3.5-Turbo-16k (Bai et al., 2024).

## 3 Motivations

The tasks of Information Retrieval and Question Answering are very related. A task of extraction is always triggered by the need to answer a question or a problem, and to perform a good Question Answering task, the machine needs to be able to identify correct and relevant information to retrieve it.

Throughout the studies done in these areas, models have been built that provided any inputs generally questions, to answer accordingly. It has been of large benefit to the industries however for the scientific community, these visions and approaches have always created new challenges. The main one is how to really identify problems among a set of information or in other words the discovery of new knowledge.

As Claude Lévi-Strauss observed, scientific advancement depends not merely on finding answers but on asking the right questions. True scientific inquiry requires machines to emulate human intelligence by formulating meaningful questions and identifying relevant problems.

Our approach is motivated by the belief that Question Generation, rather than Question Answering, represents the next step in knowledge compression and **scientific discovery systems**. We propose that documents can be more effectively segmented and understood through **question cues**, strategic points that connect related content and reveal a focus of a document without requiring complete reading. These cues serve as semantic boundaries that enhance information retrieval while facilitating a deeper understanding of document content and at the same time represent knowledge compressors that help us reduce the amount of memory that Traditional techniques cannot do in areas such as RAGs.

Our current study is part of a research series investigating how machines can generate high-quality questions that facilitate ongoing scientific discovery. Rather than directly optimizing for question answering, we explore whether generating high-quality-oriented questions and queries can indirectly enhance the retrieval capabilities of RAG systems. Without forgetting to observe its advantages and limits.

## 4 Task definition

### 4.1 Single Hop document retrieval

The objective of this task is **single document retrieval** given a user query. Within the scope of this study, the system must retrieve relevant paper IDs corresponding to the input query, without requiring any fine-tuning of the selected model. Among the retrieved IDs, only one should be an exact match to the input query.

The task can be formally defined as follows:

$$\text{Input Query} \rightarrow [Topk\_ids] \quad (1)$$

**Critical Constraint:** Only **one ID** represents a relevant or exact match within the returned set. This restriction ensures that the model functions purely as a retrieval system, returning paper IDs without generating or summarizing textual content.

### 4.2 Multihop Document Retrieval

This task, aligned with the LongBench QA benchmark (Bai et al., 2024), requires the system to answer a generally simple input query by reasoning over a maximum of four passages within a single document. In other words, it constitutes a four-hop retrieval task.

$$\text{Input Query} \rightarrow [Top\text{-}k\_answers] \quad (2)$$

## 5 Methodology

The current approach leverages well-formulated questions as effective knowledge compressors for scientific literature. We hypothesize that not only question-based representations can enhance retrieval performance but also the way by which these questions are generated plays a crucial role for obtaining performances beyond traditional chunk optimization or model fine-tuning while maintaining stable performance across diverse queries (Figure.1).

### 5.1 Main approach

The methodology consists of two main components.

**Knowledge Compression Pipeline:**

1. **Section Extraction:** We systematically extract key information-rich sections from arXiv Natural Language Processing (NLP) research

papers (methodology, introduction, discussion conclusion sections).

2. **Question Generation:** For each section, we generate both technical and conceptual ((broad) questions that capture the core contributions of papers.

3. **Queries Generation:** From generated questions and identified keywords, we construct multiple search queries associated with each paper.

4. **Card Creation:** We synthesize concise descriptive cards (limited to 300 words) containing each topic of the papers, key findings, and metadata (platform published, authors, date published, id or Doi).

5. **Embedding Computation:** Vector embeddings are computed for all questions, queries, and descriptive cards.

6. **Vector Database:** These compressed representations can be stored in a vector database, eliminating the need to maintain complete documents. However, the current approach can be implemented locally without the need of any vector Database.

**Retrieval Process:**

1. **User Query:** The system receives a natural language query from the user.

2. **Query Embedding:** This query is transformed into a vector representation.

3. **Matching:**
    - **Lexical Matching:** The query is passed through a filtering approach to match with predefined questions and queries at the lexical level.
    - **Semantic Matching:** Top questions from lexical matching undergo semantic matching and reranking through the section content from which they compress knowledge, enabling retrieval of relevant document IDs representing papers.

4. **Retrieval:** The system returns the IDs of relevant papers to the user.

The syntactic reranking used to boost the semantic retrieval of models in the second evaluation is as followed:

### 5.2 Syntactic Algorithm Description

The algorithm implements a keyword-based passage filtering and ranking system that maintains semantic coherence through order preservation:

1. **Keyword Extraction:** The query is processed using part-of-speech tagging to extract meaningful keywords by excluding adpositions (ADP), coordinating conjunctions (CCONJ), and specific punctuation marks (commas and hyphens) in order with high priority on ADP and CCONJ. This removes stop-word-like elements while preserving content-bearing terms.

2. **Frequency Scoring:** Each passage receives a score equal to the total frequency of all query keywords appearing within it. This captures both keyword coverage and repetition importance.

3. **Threshold Filtering:** Only passages with keyword frequency scores meeting or exceeding the parameter *k* are retained. This ensures a minimum level of query relevance.

4. **Order Preservation:** Rather than ranking by frequency scores, filtered passages are re-ordered according to their original sequence. This maintains the semantic and logical flow of the document collection, which is crucial for coherent information retrieval.

5. **Top-*k* Selection:** The final step returns the first 6 passages from the order-preserved filtered set, balancing relevance with manageable output size.

---

**Algorithm 1** Syntactic Reranking

---
1: **procedure** MAIN ALGORITHM(*query, passages, k*)
2:    *keywords* ← ExtractKeywordsByPOS(*query*)
3:    *analysis* ← AnalyzePassages(*keywords, passages*)
4:    *filteredPassages* ← GetFilteredPassages(*analysis, k,* 6, TRUE)
5:    **return** *filteredPassages*
6: **end procedure**

---

### 5.3 Data

The first set of experiments focuses on direct document retrieval based on an input query, using a dataset of 109 scientific papers. These papers were randomly selected from the ArXiv platform, specifically within the field of Natural Language Processing. All retrieval techniques are evaluated on this common dataset (see Table 1).

| Technique | Number of Papers |
|---|---|
| `Fixed-sized Chunking` | 109 |
| `Recursive Chunking` | 109 |
| `BM25` | 109 |

Table 1: Number of scientific papers used for each retrieval technique.

To conduct the second task (Multi-hop retrieval), we selected the LongBench QA v1 dataset, specifically `2WikiMultihopQA` and `2WikiMultihopQA_e` subsets, which contain 200 and 300[1] document rows for testing, respectively. Notably, the `2WikiMultihopQA_e` subset includes an additional 100 documents (Bai et al., 2023).

| Models | Samples of both dataset |
|---|---|
| `Llama2-7Bchat-4k` | 200-200 |
| `Vicuna-7B` | 200-200 |
| `Llama3-8B-8k` | 200-200 |

Table 2: Data distribution for the model evaluation on the datasets.

On average, each document row contains approximately 6,146 words in the `2WikiMultihopQA_e` subset and 4,887 words in the `2WikiMultihopQA` subset.

## 6 Experiments

### 6.1 Evaluation Methodology

In the first task, since only one ID constitutes a correct answer within the top-k list of results, the traditional **F1-score** metric becomes inappropriate for this task. Instead, we employ **Accuracy@k or Recall@k** as our primary evaluation metric.

#### 6.1.1 Rationale for Metric Selection

- **Single Correct Answer:** With only one relevant document per query, precision becomes less informative since it measures the proportion of retrieved documents that are relevant. In our single-answer scenario, precision will always be low (**e.g., 1/k for top-k retrieval**) regardless of retrieval quality, making it an inadequate metric for evaluation.

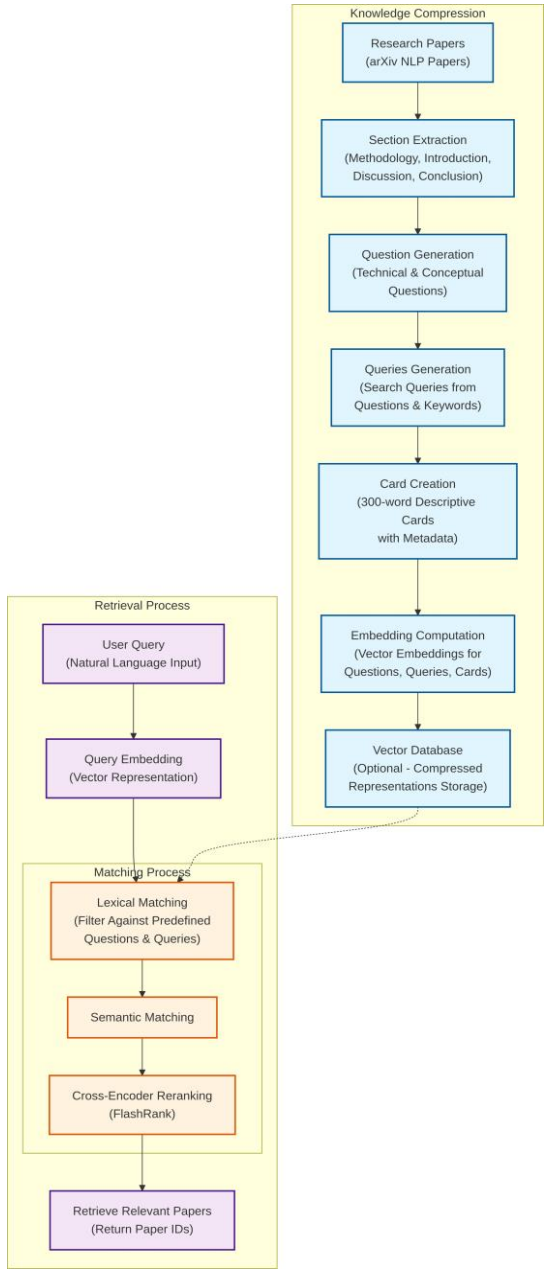

Figure 1: Knowledge Compression and Retrieval Workflow

---
[1]For the current study, the first 200 documents were selected for the 2WikiMultihopQA_e

- **Accuracy@k or Recall@k Suitability:** Both metrics in this case will present the same values and effectively capture whether the correct document appears within the top-k retrieved results, providing a binary assessment of retrieval success that aligns with our single-answer constraint.

**Additional Evaluation Metrics:** To provide a comprehensive assessment, we supplement Recall@k with additional metrics including **Mean Reciprocal Rank (MRR)**, which accounts for the position of the correct answer within the ranked list, giving higher scores to correct documents retrieved at earlier ranks.

#### 6.1.2 Evaluation Structure

We conducted two main evaluations in this task:

- **Comparison with Traditional Chunking Methods:**
    1. We evaluated our current approach against traditional chunking methods, including:
        - Structure-based retrieval
        - Fixed-size chunking
        - BM25 retrieval
    2. We also compared the effectiveness of our generated **paper-cards** with traditional abstracts in direct document retrieval tasks.

- **Benchmarking Against State-of-the-Art Methods:**
    1. We evaluated our method against state-of-the-art techniques on the same datasets used in the LongbenchQA v1 benchmark.

### 6.2 Models

The model used to construct the proposed approach and to evaluate it against traditional techniques such as BM25, recursive chunking, and fixed-size chunking is `Llama 3.2 3B Instruct`[2] (lla, 2024).

For the second evaluation, which focuses on model comparison, we employ two generative models previously used in earlier experiments, in addition to `Llama 3`. These models are:

---
[2]This model is a fine-tuned version designed for assistance and educational purposes, developed by Tethys Research. However, it is not used in the second evaluation in order to assess the approach in a **transparent** manner.

- `Llama2-7Bchat-4k` (Touvron et al., 2023),
- `Vicuna-7B` (16K context) (Zheng et al., 2023),
- `Llama3-8B` (8K context) (Dubey et al., 2024).

### 6.3 Experimental Setup

**Task 1: Single-Hop Retrieval**

In Task 1, we assess the performance of our question-centric approach against traditional retrieval methods, including fixed-size chunking, recursive chunking, and BM25. The evaluation is conducted along two dimensions:

- **Technical Queries or V2:** Technical and precise queries with strong syntactic structures.
- **Conceptual Queries or V3:** Broader, more conceptual queries requiring semantic understanding.

The queries were generated using a zero-shot prompting strategy with Claude 4, based on predefined questions and query patterns extracted from each paper (prompt details available in Appendix 10).

We also evaluate the semantic quality of the generated **paper-cards** in comparison to traditional abstracts. Both aim to preserve the essential content of a scientific paper. This comparison is carried out by measuring:

- BM25 performance using abstracts.
- BM25 performance boosted with our question-centric paper-card generation.

The retrieval model used for this task is `Llama3.2 3B Instruct`, deployed with its default configuration, except for an increased context length of 6000 tokens, for handling selected sections of papers.

**Task 2: Multi-Hop Retrieval**

For Task 2, we adopt the F1-score as the primary evaluation metric, following the LongBenchQA v1 benchmark.

We test our method using three models cited earlier in the Models section.

All models share the same inference configuration:

- Prompt templates are consistent with those used in the LongBenchQA v1 benchmark (Bai et al., 2024).

- Temperature is fixed at 0.5.

Additionally, we experiment with the syntactic reranker using two values for its main parameter L (2 and 3), which represents the minimum number of query words that must co-occur in retrieved passages independently of their position within each passage.

### 6.4 Hardware Configuration

All experiments were conducted on a system equipped with an NVIDIA GeForce RTX GPU (6GB VRAM) and 64GB RAM. We deployed models using Ollama to facilitate efficient inferences.

## 7 Results

### 7.1 Task 1

#### 7.1.1 Global performances

Table 3 presents the average performances of each traditional techniques against the question-centric approach developped.

| Approach | Acc. | MRR |
|---|---|---|
| Our Approach | **0.8440** | **0.8031** |
| Recursive Chunking | 0.2569 | 0.2173 |
| Fixed-size Chunking | 0.2317 | 0.1983 |
| BM25 | 0.7890 | 0.6770 |

Table 3: Average performance metrics across retrieval approaches.

Furthermore, the evaluations have been conducted on two dimensions (technical and conceptual queries) to evaluate the performances of the techniques at a granular level based on the nature of the queries. (see Tables 4, 5).

| Technical | @3 | | @5 | |
|---|---|---|---|---|
| | Acc. | MRR | Acc. | MRR |
| Approach | 0.8807 | **0.8578** | 0.9174 | **0.8661** |
| Recursive | 0.1651 | 0.1483 | 0.1835 | 0.1529 |
| Chunk | 0.1468 | 0.1315 | 0.1651 | 0.1352 |
| BM25 | **0.8991** | 0.8486 | **0.9358** | 0.8569 |

Table 4: Accuracy and MRR for technical or V2 queries at top-3 and top-5 retrieval.

| Conceptual | @3 | | @5 | |
|---|---|---|---|---|
| | Acc. | MRR | Acc. | MRR |
| Approach | **0.7706** | **0.7401** | **0.8073** | **0.7483** |
| Recursive | 0.3211 | 0.2798 | 0.3578 | 0.2881 |
| Chunk | 0.2936 | 0.2599 | 0.3211 | 0.2664 |
| BM25 | 0.6055 | 0.4893 | 0.7156 | 0.5131 |

Table 5: Accuracy and MRR for conceptual or V3 queries at top-3 and top-5 retrieval.

#### 7.1.2 BM25 Boosted

We evaluate the performances of BM25 on both abstract papers and the paper-card generated from our approach. The results can be seen on Table (6, 7).

Figure 3 presents a synthesis of the performances of BM25 on both techniques.

| Technical | @3 | | @5 | |
|---|---|---|---|---|
| | Acc. | MRR | Acc. | MRR |
| BM25 (Card) | **0.9633** | **0.9373** | **0.9725** | **0.9396** |
| BM25 (Abs) | 0.8991 | 0.8456 | 0.9358 | 0.8543 |

Table 6: Accuracy and MRR for BM25 technical queries (v2) at top-3 and top-5 retrieval.

| Conceptual | @3 | | @5 | |
|---|---|---|---|---|
| | Acc. | MRR | Acc. | MRR |
| BM25 (Card) | **0.8899** | **0.8502** | **0.9174** | **0.8557** |
| BM25 (Abs) | 0.6606 | 0.5627 | 0.7156 | 0.5751 |

Table 7: Accuracy and MRR for BM25 conceptual queries (v3) at top-3 and top-5 retrieval.

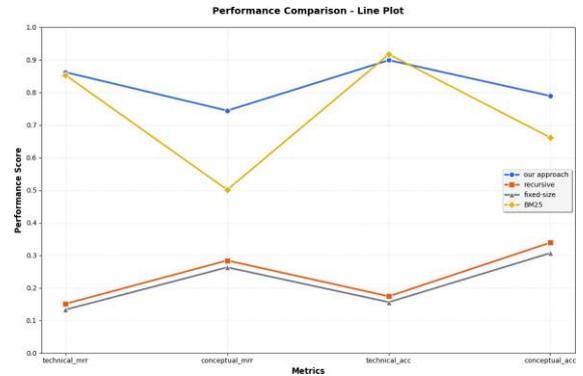

Figure 2: Global perfomances of the techniques.

### 7.2 Task 2

Table 8 presents the evaluation on the second task concerning the multihop retrieval performances of the models on the selected benchmark. The first step is to study the direct performances of the models on the 2WikiMultihopQA dataset and compare the results against the other implementations mostly using finetuning techniques.

| Models | F1 |
|---|---|
| **Llama2-7B-chat-4k (ours)** | **0.520** |
| CFIC-7B (Llama2-7B-chat improved) | 0.412 |
| Llama2-7B-chat-4k (Bai et al., ACL 2024) | 0.328 |
| **Vicuna-7B (ours)** | **0.340** |
| Vicuna-7B (CFIC, Yang et al. 2024) | 0.233 |
| **Llama3 (ours)** | **0.550** |
| GPT-3.5-Turbo-16k (Bai et al., ACL 2024) | 0.377 |
| Llama_index (Zhao et al., 2024) | 0.117 |
| PPL chunking (Zhao et al., 2024, Qwen2) | 0.141 |

Table 8: Performance comparison across models.

Then we evaluated `Llama3` and `Llama2-7B-chat-` on the `2WikiMultihopQA_e` to evaluate at the granular level the performances of the models from 0k-4k to 4k-8k context windows using the syntactic parameter L.

| Models | 0-4K | 4K-8K |
|---|---|---|
| Llama2-7B-chat-4k_L3(ours) | 0.470 | 0.500 |
| Llama2-7B-chat-4k_L2(ours) | 0.560 | 0.470 |
| Llama3_L3(ours) | 0.550 | **0.580** |
| Llama3_L2(ours) | **0.570** | 0.550 |
| GPT-3.5-Turbo-16k | 49.8 | 45.1 |
| Llama2-7B-chat-4k | 33.3 | 22.5 |

Table 9: F1-Score Performance Across Context Ranges on 2WikiMultihopQA_e using parameter L at 2 or 3.

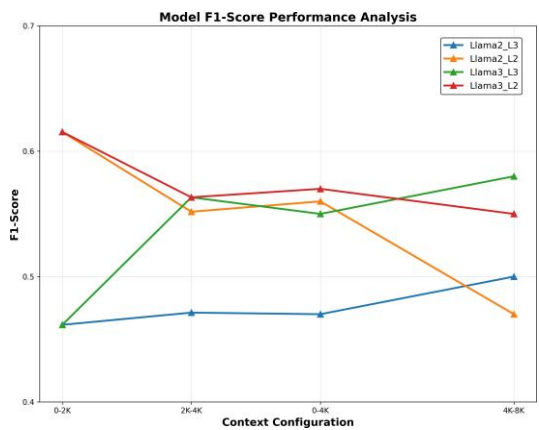

Figure 4: Perfomances on the 2WikiMultihopQA_e. LLama3 and LLama2 correspond to the models using our approach with some specificities on the parameter L which can be either 3 or 2.

## 8 Discussions

Our investigation began with the hypothesis that queries and question anchors enhance information retrieval effectiveness in document-based RAG systems

### 8.1 Task 1

The first evaluation conclusively demonstrates that our approach outperforms traditional chunking techniques when applied to scientific research papers in Natural Language Processing. We observe in Table 3 that our approach generally performs best across the dataset. However, when examining performance at a more granular level based on the technicity of the query, BM25 achieves higher Accuracy on technical queries, while our method yields better results on Mean Reciprocal Rank (MRR). This can be explained by the fact that BM25 is strongly syntactic. Technical queries tend to be more precise and keyword-driven, which enables BM25 to retrieve relevant documents more effectively. However, it performs poorly on conceptual or broader queries, which require deeper semantic understanding to relate topics and documents.

As shown in Figure 2, our approach demonstrates greater stability than BM25 across both technical and conceptual queries. BM25 shows a sharp drop in MRR from 84 percent to 48 percent, almost a 50 percent decline, whereas our question-centric approach exhibits only a modest decrease of approximately 12 percent, highlighting the instability of BM25 on semantically complex queries.

On the other hand, fixed-size and recursive chunking techniques struggle to exceed 40 percent Accuracy. This is likely due to the inherent complexity of scientific papers. Interestingly, these methods tend to perform better on conceptual queries than on technical ones.

A deeper analysis of BM25 performance in Figure 3 shows that combining the paper cards generated by our approach not only boosts BM25 performance but also significantly improves its stability compared to using traditional paper abstracts.

The MRR and Accuracy scores across both query types remain stable and show a gradual increase when moving from top-3 to top-5 retrieved answers.

Figure 5 presents a comparative view of the metrics where our approach demonstrates stronger performance.

Finally, the approach offers another advantage. Its performance goes beyond Exact Match and MRR metrics, extending to significant benefits in computational efficiency and storage requirements. During the experiments, our method consistently demonstrated substantially lower storage demands

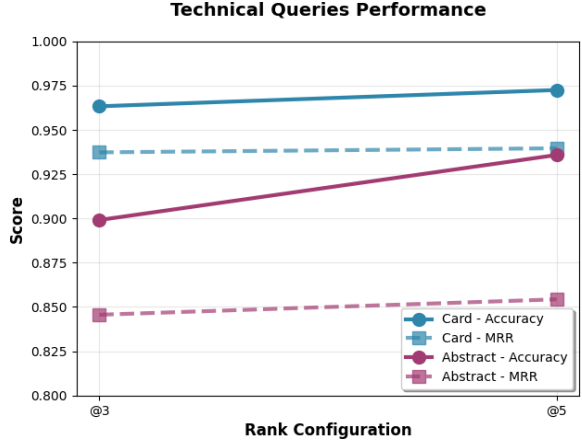
(a) BM25 performance on technical queries (v2).

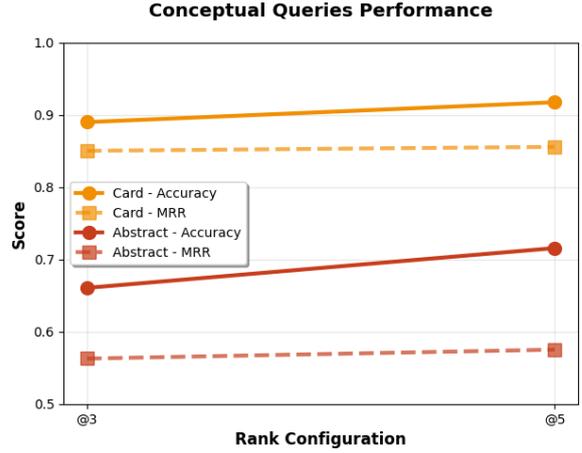
(b) BM25 performance on conceptual queries (v3).

Figure 3: Comparison of BM25 performance across technical (v2) and conceptual (v3) queries.

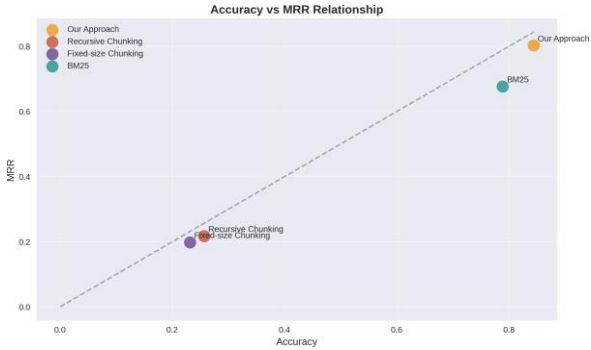
(a) Gap performances between approaches.

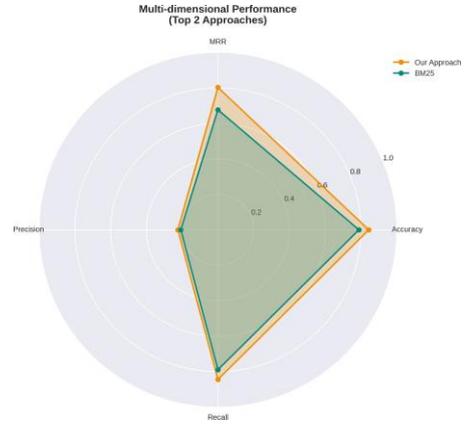
(b) Performance comparison between BM25 and our approach.

Figure 5: Gap performance comparison between techniques on Accuracy, MRR and recall

compared to conventional techniques.

For a corpus of 109 papers, our approach requires storing only two indexes containing approximately 218 embedding vectors (109 embeddings per index). In contrast, recursive chunking techniques produced 38,509 chunks, while fixed-size chunking generated 44,809 chunks for the same dataset (Table 10).

| Methods | Number of Records |
| --- | --- |
| Recursive Chunking | 38,509 |
| Fixed-size Chunking (350 words) | 44,809 |
| paper-card (ours) | **109** |
| main-research-queries (ours) | **95** |

Table 10: Sotrage efficiency comparison

Storage efficiency is further demonstrated by the compact size of our paper-cards, which are stored in Markdown format and rarely exceed 5KB (with the largest being only 3.1KB).

## 8.2 Task 2

As shown in Table 8, applying the developed approach to the models leads to better performance than previous implementations, particularly when using the hyperparameter $L$ fixed at 3 (see Table 9).

In Figure 4, we observe that models configured with the hyperparameter set to 3 tend to improve in performance as the context window increases. In contrast, models using $L = 2$ initially achieve high scores but experience a performance drop as more context is introduced, likely due to the accumulation of textual noise.

Additionally, we note that Llama 3 and `Llama2-7B-Chat-4k`, when enhanced with the pro-posed approach, show only a marginal difference in F1 score, approximately 3%. Furthermore, there is a notable performance gain of around 20% between `Llama 2-7B-Chat-4k` (Bai et al., 2023) in its original form and our enhanced version. This suggests that a significant portion of the observed improvements can be attributed to enhanced syntactic understanding.

### 8.3 Advantages of the proposed approach

Our approach offers multiple advantages:

- **Storage Efficiency:** Dramatically reduced storage requirements compared to traditional chunking methods.

- **Implementation Simplicity:** No need for model fine-tuning, unlike approaches evaluated in our second experiment, our models remained untuned for chunking tasks.

- **Inference Speed:** Near-instantaneous retrieval as the system matches queries (few words) and reranks paper-cards (maximum 300 characters), eliminating the lengthy processing times associated with on-demand query generation approaches like RAG-Fusion proposed by Zackary R.

- **Semantic Preservation:** Maintains semantic richness without storing entire documents.

- **Intuitive Access:** Supports more natural information retrieval through question-based retrieval with both local and remote matching capabilities.

## 9 Conclusion and Future Work

This study investigated whether query-oriented and question generation, used as a form of knowledge compression, can improve retrieval performance in Retrieval-Augmented Generation (RAG) systems while maintaining semantic stability. The proposed approach significantly outperformed traditional chunking methods and BM25, providing greater robustness and consistency. When used in combination with BM25, our method further enhanced retrieval quality, indicating complementary strengths.

In multi-hop retrieval, we demonstrated that a granular syntactic reranking strategy enables models to identify and retrieve relevant passages for more accurate answers. These results were validated through the `LongBench QA v1` benchmark, with particularly strong performance on the `2WikiMultihopQA` and `2WikiMultihopQA_e` subsets.

Several limitations were identified during experimentation, which point to future areas of improvement. As part of our future work, we plan to expand our evaluation to include additional benchmark datasets such as:

- `LongBench QA v2`
- `Qasper`
- `FinanceBenchQA`
- `NFcorpus`
- `Loong`

We also intend to introduce semantic evaluation metrics, including BLEU, ROUGE, and other relevant measures, to better assess the quality of both retrieval and generation. Furthermore, we aim to develop a structured graph-based representation of the generated queries and questions, which may enhance interpretability and improve performance in complex retrieval scenarios.

Addressing these directions will help consolidate our approach and strengthen its applicability in real-world long-context retrieval and generation systems.

## 10 Limitations

During the experiments, several limitations were identified. From the model side, we observed challenges in generating concise and focused questions or queries, especially when dealing with long contexts. This was particularly evident in the verbose generation of keywords and the lack of in-depth questions in the multi-hop retrieval task, where passages could exceed 10,000 words.

However, the use of a syntactic reranker helped in curating more relevant passages, demonstrating its utility in improving retrieval quality.

These observations suggest that current state of the art models are not yet sufficiently trained to generate high quality questions ranging from simple to complex given complex and long passages. This shortcoming may stem, in part, from limited syntactic understanding. While semantic comprehension is important, crafting effective questions also requires robust and stable syntactic awareness, particularly regarding token relationships. A stronger grasp of syntax could enable models to better identify critical keywords and decompose complex questions into simpler sub-questions.

Another limitation of this research lies in the organization of the generated questions. Although the questions are linked to specific research papers or documents, there is no broader structural organization that facilitates retrieval as the dataset scales. Structuring these questions and their sources into a graph-based representation could offer a more robust and scalable solution for fast and high quality retrieval across large corpora.

We also observed that the absence of fine tuning posed a significant limitation for retrieval performance. In particular, syntactic fine tuning could offer more substantial improvements than approaches focusing solely on semantic fine tuning, which generally bias the model toward specific domain areas. Fine tuning on syntactic structures may help models perform more effectively generally across domains.

Finally, a key limitation of the syntactic reranker lies in its dependency on a fixed parameter $L$. The current design does not allow the reranker to adaptively determine the optimal value for $L$; instead, it remains fixed. A meaningful improvement would be to extend the reranker logic to operate over a dynamic range of values for $L$, enabling it to automatically adjust based on context and passage relevance.

## Acknowledgements

The work was done with partial support from the Mexican Government through the grant A1- S-47854 of CONACYT, Mexico, grants 20250738, 20241819, and 20240951 of the Secretaría de Investigación y Posgrado of the Instituto Politécnico Nacional, Mexico. The authors thank the CONACYT for the computing resources brought to them through the Plataforma de Aprendizaje Profundo para Tecnologías del Lenguaje of Unnumbered acknowledgements section if required. the Laboratorio de Supercómputo of the INAOE, Mexico, and acknowledge the support of Microsoft through the Microsoft Latin America PhD Award.

## References

2024. Llama 3.2: Revolutionizing edge AI and vision with open, customizable models.

Yushi Bai, Xin Lv, Jiajie Zhang, Hongchang Lyu, Jiankai Tang, Zhidian Huang, Zhengxiao Du, Xiao Liu, Aohan Zeng, Lei Hou, Yuxiao Dong, Jie Tang, and Juanzi Li. 2023. Longbench: A bilingual, multi-task benchmark for long context understanding.

Yushi Bai, Xin Lv, Jiajie Zhang, Hongchang Lyu, Jiankai Tang, Zhidian Huang, Zhengxiao Du, Xiao Liu, Aohan Zeng, Lei Hou, Yuxiao Dong, Jie Tang, and Juanzi Li. 2024. LongBench: A bilingual, multi-task benchmark for long context understanding. In *Proceedings of the 62nd Annual Meeting of the Association for Computational Linguistics (Volume 1: Long Papers)*, pages 3119–3137, Bangkok, Thailand. Association for Computational Linguistics.

Jacob Devlin, Ming-Wei Chang, Kenton Lee, and Kristina Toutanova. 2019. BERT: Pre-training of deep bidirectional transformers for language understanding. In *Proceedings of the 2019 Conference of the North American Chapter of the Association for Computational Linguistics: Human Language Technologies, Volume 1 (Long and Short Papers)*, pages 4171–4186, Minneapolis, Minnesota. Association for Computational Linguistics.

André V. Duarte, João Marques, Miguel Graça, Miguel Fernández Freire, Lei Li, and Arlindo L. Oliveira. 2024. Lumberchunker: Long-form narrative document segmentation. *ArXiv*, abs/2406.17526.

Abhimanyu Dubey, Abhinav Jauhri, Abhinav Pandey, Abhishek Kadian, Ahmad Al-Dahle, Aiesha Letman, Akhil Mathur, Alan Schelten, Amy Yang, Angela Fan, Anirudh Goyal, Anthony S. Hartshorn, Aobo Yang, Archi Mitra, Archie Sravankumar, Artem Korenev, Arthur Hinsvark, Arun Rao, Aston Zhang, Aur'elien Rodriguez, Austen Gregerson, Ava Spataru, Baptiste Rozière, Bethany Biron, Binh Tang, Bobbie Chern, Charlotte Caucheteux, Chaya Nayak, Chloe Bi, Chris Marra, Chris McConnell, Christian Keller, Christophe Touret, Chunyang Wu, Corinne Wong, Cris tian Cantón Ferrer, Cyrus Nikolaidis, Damien Allonsius, Daniel Song, Danielle Pintz, and ... 2024. The llama 3 herd of models. *ArXiv*, abs/2407.21783.

Hongyu Gong, Yelong Shen, Dian Yu, Jianshu Chen, and Dong Yu. 2020. Recurrent chunking mechanisms for long-text machine reading comprehension. In *Proceedings of the 58th Annual Meeting of the Association for Computational Linguistics*, pages 6751–6761, Online. Association for Computational Linguistics.

Antonio Jimeno-Yepes, Yao You, Jan Milczek, Sebastian Laverde, and Ren-Yu Li. 2024. Financial report chunking for effective retrieval augmented generation. *ArXiv*, abs/2402.05131.

Patrick Lewis, Ethan Perez, Aleksandara Piktus, Fabio Petroni, Vladimir Karpukhin, Naman Goyal, Heinrich Kuttler, Mike Lewis, Wen tau Yih, Tim Rocktäschel, Sebastian Riedel, and Douwe Kiela. 2020. Retrieval-augmented generation for knowledge-intensive nlp tasks. *ArXiv*, abs/2005.11401.

Carlo Merola and Jaspinder Singh. 2025. Reconstructing context: Evaluating advanced chunking strategies for retrieval-augmented generation.

Hongjin Qian, Zheng Liu, Kelong Mao, Yujia Zhou, and Zhicheng Dou. 2024. Grounding language model with chunking-free in-context retrieval. In *Proceedings of the 62nd Annual Meeting of the Association for Computational Linguistics (Volume 1: Long Papers)*, pages 1298–1311, Bangkok, Thailand. Association for Computational Linguistics.

Renyi Qu, Ruixuan Tu, and Forrest Sheng Bao. 2025. Is semantic chunking worth the computational cost? In *Findings of the Association for Computational Linguistics: NAACL 2025*, pages 2155–2177, Albuquerque, New Mexico. Association for Computational Linguistics.

Zackary Rackauckas. 2024. Rag-fusion: a new take on retrieval-augmented generation. *ArXiv*, abs/2402.03367.

Hugo Touvron, Louis Martin, Kevin R. Stone, Peter Albert, Amjad Almahairi, Yasmine Babaei, Nikolay Bashlykov, Soumya Batra, Prajjwal Bhargava, Shruti Bhosale, Daniel M. Bikel, Lukas Blecher, Cristian Cantón Ferrer, Moya Chen, Guillem Cucurull, David Esiobu, Jude Fernandes, Jeremy Fu, Wenyin Fu, Brian Fuller, Cynthia Gao, Vedanuj Goswami, Naman Goyal, Anthony S. Hartshorn, Saghar Hosseini, Rui Hou, Hakan Inan, Marcin Kardas, Viktor Kerkez, Madian Khabsa, Isabel M. Kloumann, Artem Korenev, Punit Singh Koura, Marie-Anne Lachaux, Thibaut Lavril, Jenya Lee, Diana Liskovich, Yinghai Lu, Yuning Mao, Xavier Martinet, Todor Mihaylov, Pushkar Mishra, Igor Molybog, Yixin Nie, Andrew Poulton, Jeremy Reizenstein, Rashi Rungta, Kalyan Saladi, Alan Schelten, Ruan Silva, Eric Michael Smith, R. Subramanian, Xia Tan, Binh Tang, Ross Taylor, Adina Williams, Jian Xiang Kuan, Puxin Xu, Zhengxu Yan, Iliyan Zarov, Yuchen Zhang, Angela Fan, Melissa Hall Melanie Kambadur, Sharan Narang, Aur'elien Rodriguez, Robert Stojnic, Sergey Edunov, and Thomas Scialom. 2023. Llama 2: Open foundation and fine-tuned chat models. *ArXiv*, abs/2307.09288.

Shicheng Xu, Liang Pang, Huawei Shen, and Xueqi Cheng. 2023. BERM: Training the balanced and extractable representation for matching to improve generalization ability of dense retrieval. In *Proceedings of the 61st Annual Meeting of the Association for Computational Linguistics (Volume 1: Long Papers)*, pages 6620–6635, Toronto, Canada. Association for Computational Linguistics.

Jihao Zhao, Zhiyuan Ji, Jason Zhaoxin Fan, Hanyu Wang, Simin Niu, Bo Tang, Feiyu Xiong, and Zhiyu Li. 2025. Moc: Mixtures of text chunking learners for retrieval-augmented generation system. *ArXiv*, abs/2503.09600.

Jihao Zhao, Zhiyuan Ji, Pengnian Qi, Simin Niu, Bo Tang, Feiyu Xiong, and Zhiyu Li. 2024. Meta-chunking: Learning efficient text segmentation via logical perception. *arXiv preprint arXiv:2410.12788*.

Lianmin Zheng, Wei-Lin Chiang, Ying Sheng, Siyuan Zhuang, Zhanghao Wu, Yonghao Zhuang, Zi Lin, Zhuohan Li, Dacheng Li, Eric P. Xing, Haotong Zhang, Joseph E. Gonzalez, and Ion Stoica. 2023. Judging llm-as-a-judge with mt-bench and chatbot arena. *ArXiv*, abs/2306.05685.

Zhiying Zhu, Zhiqing Sun, and Yiming Yang. 2024. Halueval-wild: Evaluating hallucinations of language models in the wild. *ArXiv*, abs/2403.04307.

# Appendix

**Prompts used to generate queries** from predefined questions and queries:

> Generate only one query for each, which combines the main technical approach or method and the application domain or specific technique.

> generate csv documents of a query per item based on the queries and questions that result in a simulated simple user query that combines the main technical approach or method and the broader goal or alternative perspective.